\documentclass{svproc}
\usepackage{graphicx}
\usepackage{url}

\begin{document}
\mainmatter              
\title{Charged-particle decays of highly excited states in $^{19}$F.}
\titlerunning{Charged-particle decays of highly excited states in $^{19}$F.}  
%
\author{P. Adsley\inst{1} \and F. Hammache\inst{1} \and N. de S\'{e}r\'{e}ville \and M. Assi\'{e}\inst{1} \and D. Beaumel\inst{1} \and M. Chabot\inst{1} \and M. Degerlier\inst{1} \and C. Delafosse\inst{1} \and F. Flavigny\inst{1} \and A. Georgiadou\inst{1} \and J. Guillot\inst{1} \and V. Guimar\~{a}es\inst{1} \and A. Gottardo\inst{1} \and I. Matea\inst{1} \and L. Olivier\inst{1} \and L. Perrot\inst{1} \and I. Stefan\inst{1} \and A. Laird\inst{2} \and S. P. Fox\inst{2} \and R. Garg\inst{2} \and S. Gillespie\inst{2} \and J. Riley\inst{2} \and J. Kiener\inst{3} \and A. Lefebvre-Schuhl\inst{3} \and V. Tatischeff\inst{3} \and I. Sivacek\inst{4}}

\authorrunning{P. Adsley {\it et al.}} 
%
\tocauthor{P. Adsley, F. Hammache, N. de S\'{e}r\'{e}ville, M. Assi\'{e}, D. Beaumel, M. Chabot, M. Degerlier, C. Delafosse, F. Flavigny, A. Georgiadou, J. Guillot, V. Guimarã\~{a}es, A. Gottardo, I. Matea, L. Olivier, L. Perrot, I. Stefan, A. Laird, S. P. Fox, R. Garg, S. Gillespie, J. Riley, J. Kiener, A. Lefebvre-Schuhl, V. Tatischeff, I. Sivacek}
\institute{Institut Physique Nucl\'{e}aire d’Orsay, UMR8608, CNRS-IN2P3, Université Paris Sud 11, 91406 Orsay, France
\\
\email{padsley@gmail.com},\\
\and
 Department of Physics, University of York, Heslington, York, YO10 5DD, United Kingdom
\and
Centre de Sciences Nucl\'{e}aires et de Sciences de la Mati\`{e}re (CSNSM), CNRS/IN2P3,
Univ. Paris-Sud, Universit\'{e} Paris–Saclay, B\^{a}timent 104, F–91405 Orsay Campus, France
\and
ASCR-Rez, CZ-250 68, Rez, Czech Republic
}

\maketitle              

\begin{abstract}

Neutron-capture reactions on $^{18}$F in the helium-burning shell play an important role in the production of $^{15}$N during core-collapse supernovae. The competition between the $^{18}$F($n,p/\alpha$)$^{18}$O/$^{15}$N reactions controls the amount of $^{15}$N produced. The strengths of these reactions depend on the decay branching ratios of states in $^{19}$F above the neutron threshold. We report on an experiment investigating the decay branching ratios of these states in order to better constrain the strengths of the reactions.

\end{abstract}
\section{Astrophysical Background}
Spatially correlated hot-spots of $^{15}$N and $^{18}$O have been observed in grains which originate from core-collapse supernovae \cite{OrgeuilMeteorite}. In the helium-burning shell, $^{14}$N produced during the CNO cycles is converted into $^{18}$F and $^{18}$O by $^{14}$N($\alpha,\gamma$)$^{18}$F($\beta^+$)$^{18}$O. During the supernovae the $^{18}$O($\alpha,n$)$^{21}$Ne reaction activates releasing neutrons and causing $^{18}$F($n,p/\alpha$)$^{18}$O/$^{15}$N reactions. The amounts of $^{15}$N and $^{18}$O produced depend sensitively on the relative strength of the $^{18}$F($n,p/\alpha$)$^{18}$O/$^{15}$N reactions \cite{BojaziAndMeyer} determined by $^{19}$F states above the neutron threshold. At present, the astrophysical reaction rates used for astrophysical models are based on statistical-model calculations \cite{BojaziAndMeyer}; as $^{19}$F is known to exhibit strong clustering behaviour \cite{NeonFluorineClustering} statistical-model calculations may be inappropriate.

Direct measurement of the pertinent cross sections is extremely challenging as neither neutrons nor $^{18}$F may be easily fashioned into targets. Instead, the rates may be calculated from detailed knowledge of the properties (energies, total and partial widths, and spins and parities) of the nuclear states above the neutron threshold in $^{19}$F.

We report a study of excited states by $^{19}$F($p,p^\prime$) to constrain the $^{18}$F($n,p$)$^{18}$O and $^{18}$F($n,\alpha$)$^{15}$N reaction rates.

\section{Experimental Setup}
A 15-MeV proton beam from the Orsay tandem was incident upon a 90-$\mu$g/cm$^2$ LiF foil on a carbon backing located at the target position of the `Split-Pole' Enge magnetic spectrometer. Scattered particles were momentum analysed in the spectrometer and detected at the focal plane in a position-sensitive gas detector, a gas proportional detector and a plastic scintillator.

Charged particles decaying from the populated states in $^{19}$F were detected in an array of six W1 double-sided silicon strip detectors (DSSSDs). As the beamstop was located within the scattering chamber of the spectrometer, a steel shield was placed in the chamber to reduce the background seen by the silicon detectors.
\section{Data Analysis and Preliminary Results}

\begin{figure}
\includegraphics[width=\textwidth]{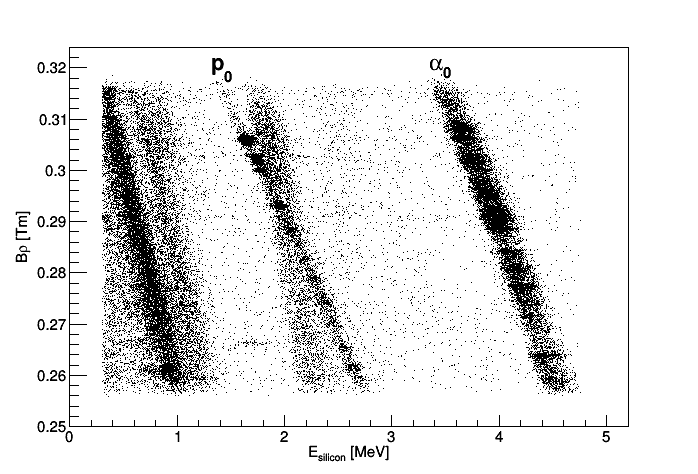}
 \caption{Magnetic rigidity against the energy detected in the silicon detectors for detector number 5. The $\alpha_0$ and $p_0$ loci are marked. The other loci are due to exited states in $^{12}$C from the target backing breaking up and competing coincidence channels such as $^{19}$F($p,\alpha$)$^{16}$O($p$) reactions.}
 \label{fig:CoincidenceSpectrum}
\end{figure}

Figure \ref{fig:CoincidenceSpectrum} shows the coincidence spectrum where there is a hit in the focal plane and a hit in a silicon detector after imposing certain conditions on the data to check for good hits in the silicon detectors .

\begin{figure}
\includegraphics[width=\textwidth]{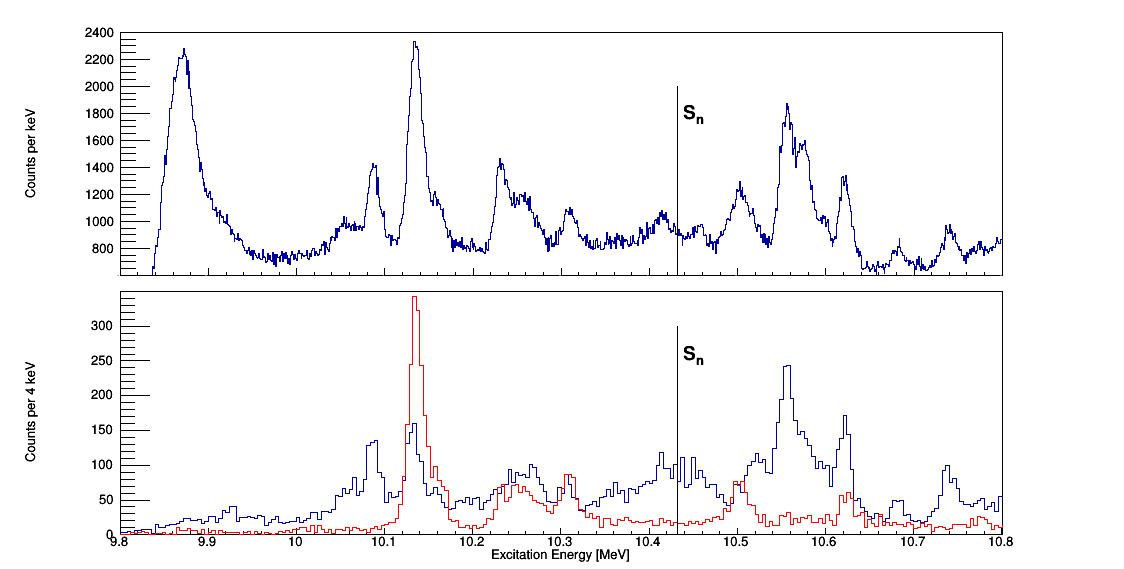}
 \caption{(Top) Singles excitation-energy spectrum. (Bottom) Excitation-energy spectra gated on $\alpha_0$ (blue) and $p_0$ (red) decays.}
 \label{fig:Spectra}
\end{figure}

The focal-plane spectra gated on a particular decay channel could then be constructed by selecting those events where the `missing' energy corresponded to the separation energy for the channel of interest. Excitation-energy spectra for the singles events, the $\alpha_0$-decay-gated events and the $p_0$-decay-gated events are shown in Figure \ref{fig:Spectra}.

Once the various spectra have been generated they are fitted to extract information on the state parameters e.g. the excitation energies, total widths and decay branches. We started by fitting the spectra using known state information taken from the ENSDF database \cite{ENSDF}. However, we found that a number of states which have been observed in resonance reaction measurements particularly of $^{18}$O($p,\alpha$)$^{15}$N \cite{Carlson,Gorodetzky1,Gorodetzky2} have been omitted by the compilers of the nuclear data-sheets \cite{Compilation} complicating the analysis as well-defined physical parameters for these states are not available and the present experiment is unable to resolve them.

The first important observation from the present experiment is the weakness of the proton decays for many of the excited states. We can compare the ratios of the observed branching ratios to the reaction rates from statistical-model calculations \cite{TALYS}. The calculations predict that the $^{18}$F($n,\alpha$)$^{15}$N reaction rate should be around three times stronger than the $^{18}$F($n,p$)$^{18}$O reaction. From the results of the present experiment, we find that the $^{18}$F($n,p$)$^{18}$O reaction rate is much weaker relative to the $^{18}$F($n,\alpha$)$^{15}$N reaction rate than that predicted using TALYS with a corresponding increase in the production of $^{15}$N in supernovae.

\section{Outlook}
More $^{19}$F($p,p^\prime$) data have been collected using the Q3D magnetic spectrometer at MLL, Garching in July 2018. The excitation-energy resolution achievable in this case is around 5 keV \cite{AdsleyMg26} which is around a factor of three better than the current experiment. This should allow for the states in the region of interest to be better resolved and used to guide the future progress of the coincidence analysis.
%

\end{document}